\documentclass[]{fairmeta}

\usepackage{float}
\usepackage{amsmath}
\usepackage{amssymb}
\usepackage{amsfonts}
\usepackage{amsthm}
\usepackage{mathrsfs}
\usepackage{makecell}
\usepackage{bbding}
\usepackage{wrapfig}
\usepackage{diagbox}
\usepackage[ruled]{algorithm2e}
\usepackage{enumerate}
\usepackage{enumitem}
\usepackage{xspace}

\titleformat*{\paragraph}{\normalsize\sffamily\bfseries}

\newcommand{\maybeincludegraphics}[2][]{%
  \IfFileExists{#2}{%
    \includegraphics[#1]{#2}%
  }{%
    \fbox{\parbox[c][2in][c]{0.9\linewidth}{\centering Missing graphic\\\texttt{\detokenize{#2}}}}%
  }%
}

\newcommand*{\system}{WavFlow\@\xspace}

\title{{\color{metablue}\system}: Audio Generation in Waveform Space}

\author[1,2]{Feiyan Zhou}
\author[1]{Luyuan Wang}
\author[1,\text{$\ast$}]{Shoufa Chen}
\author[1]{Zhe Wang}
\author[1]{Zhiheng Liu}
\author[1]{Yuren Cong}
\author[1]{Xiaohui Zhang}
\author[1]{Fanny Yang}
\author[1]{Belinda Zeng}
\affiliation[1]{Meta AI}
\affiliation[2]{Northeastern University}
\contribution[\text{$^\ast$}]{Corresponding author}

\date{\today}

\metadata[Project page]{\url{https://facebookresearch.github.io/WavFlow/}}
\metadata[Code]{\url{https://github.com/facebookresearch/WavFlow}}

\abstract{Modern audio generation predominantly relies on latent-space compression, introducing additional complexity and potential information loss. In this work, we challenge this paradigm with WavFlow, a framework that generates high-fidelity audio directly in raw waveform space without intermediate representations. To overcome the inherent difficulties of modeling high-dimensional and low-energy signals, we reshape audio into 2D token grids through waveform patchify and introduce amplitude lifting to align signal scales, enabling stable optimization via direct $x$-prediction in flow matching. To capture complex semantic alignment and temporal synchronization, we leverage an automated data pipeline to curate 5\,M high-quality video-text-audio triplets, allowing the model to learn fine-grained acoustic patterns from scratch. Experimental results show that WavFlow achieves competitive results on the video-to-audio benchmark VGGSound (FD$_{\text{PaSST}}$ 59.98, IS$_{\text{PANNs}}$ 17.40, DeSync 0.44) and the text-to-audio benchmark AudioCaps (FD$_{\text{PANNs}}$ 10.63, IS$_{\text{PANNs}}$ 12.62), matching or exceeding the performance of established latent-based methods. Our work demonstrates that such intermediate compression is not a prerequisite for high-quality synthesis, offering a simpler and more scalable alternative for multimodal audio generation.}

\begin{document}
\maketitle

\section{Introduction}
\label{sec:intro}

\begin{figure}[!t]
\centering
\maybeincludegraphics[width=0.80\linewidth]{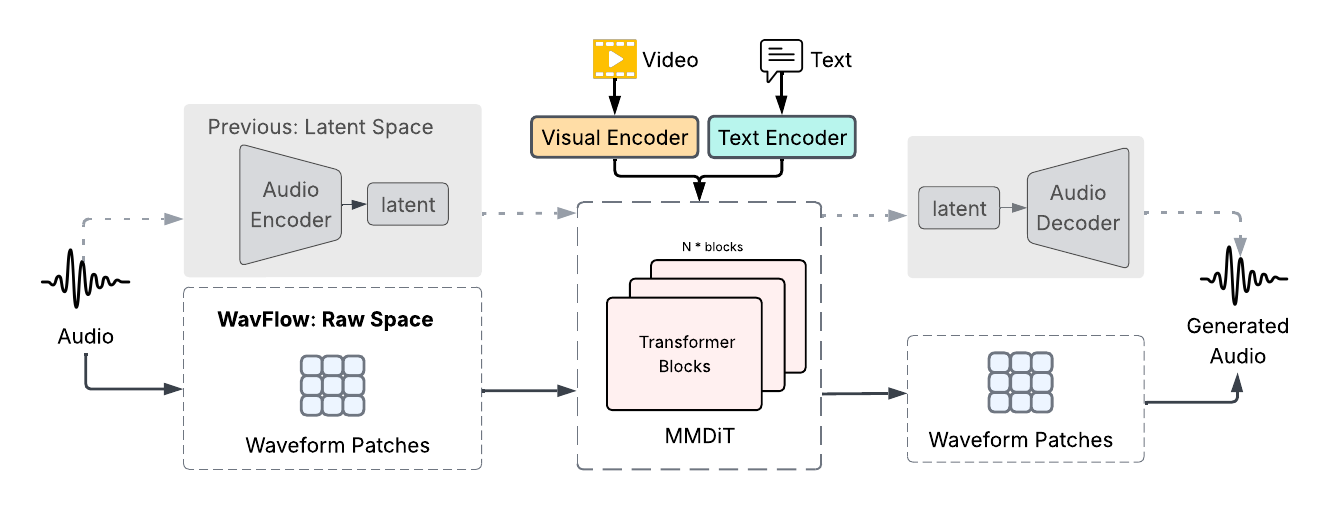}
\vspace{-0.1in}
\caption{\textbf{Standard Latent-Space vs. WavFlow.} WavFlow eliminates the encoding-decoding bottleneck by processing waveform patches directly in the raw space.}
\label{fig:composite}
\end{figure}

Video-to-audio synthesis, often referred to as Foley-style generation~\citep{mmaudio,foleycrafter,hunyuanfoley,klingfoley}, aims to produce environmental and event-based soundscapes temporally and semantically aligned with the visual content. Recent state-of-the-art methods~\citep{polyak2025moviegencastmedia,luo2024diff,foleycrafter,frieren,mmaudio,hunyuanfoley,thinksound,liu2025prismaudiodecomposedchainofthoughtsmultidimensional,klingfoley,dai2026omni2soundunifiedvideotexttoaudiogeneration,tian2025audiox} have made rapid progress by adopting a common latent-space recipe: raw signals are first mapped into a compressed representation by a pretrained tokenizer or VAE~\citep{defossez2022high,zeghidour2021soundstream,kumar2024high,evans2024fasttimingconditionedlatentaudio,kong2020hifi,lee2023bigvgan}, then a multimodal diffusion or flow-matching transformer~\citep{ho2020denoising,rombach2022high,lipman2023flow,sd3,dit} learns their conditional distribution given visual and text features. Finally, a decoder reconstructs the waveform from these generated latents, as shown in Figure~\ref{fig:composite} (top). This paradigm has become the dominant framework for modern audio generation tasks.

While effective, this approach leaves a foundational question open: is latent-space compression truly necessary for audio generation? Relying on a separate, pretrained tokenizer not only increases pipeline complexity but also constrains the final synthesis quality to the reconstruction fidelity. This motivates us to investigate direct raw-waveform generation as a way to achieve strict temporal and semantic alignment while bypassing the intermediate compression layer.

Doing so, however, is non-trivial, as raw audio differs from latent representations in three fundamental ways. First, raw waveforms are extremely high-dimensional, leading to long sequences that are computationally challenging to model directly. Second, waveform amplitudes exhibit a high dynamic range while heavily concentrating near zero, yielding a poor signal-to-noise ratio during training that makes the flow-matching objective difficult to optimize in raw space.
Third, paired video-audio datasets remain relatively scarce. Even the widely-used VGGSound~\citep{chen2020vggsound} contains only $\sim$200K samples (500 hours), a scale insufficient for models that operate directly on raw waveforms, which must learn complex acoustic structures, temporal dynamics, and precise cross-modal alignments end-to-end without the inductive bias provided by encoded audio priors.

In this work, we introduce \system, a generative framework that performs Foley-style audio synthesis directly in raw waveform space. The architecture is deliberately simple: as illustrated in Figure~\ref{fig:composite} (bottom), we employ \emph{waveform patchify} to reshape high-dimensional 1D waveforms into 2D token grids, and adopt $x$-prediction~\citep{li2026basicsdenoise} under conditional flow matching as a more stable training target for raw signals. To bridge the signal intensity mismatch between raw waveforms and the unit-variance Gaussian prior, we incorporate RMS normalization and amplitude scaling, lifting the signal into a range conducive to generative modeling. Finally, to address the data scarcity in raw-space learning, we develop an automated curation pipeline to filter a large-scale media data for audio quality and event diversity, yielding approximately $5$\,M high-quality video-text-audio pairs~\citep{polyak2025moviegencastmedia}. We train \system on this curated dataset to achieve robust video-conditioned generation and extend the model to text-only audio generation by simply zeroing out the visual conditions.

We evaluate \system on the standard VT2A (VGGSound) and T2A (AudioCaps~\citep{kim2019audiocaps}) benchmarks. On VGGSound, \system achieves state-of-the-art FD$_{\text{PaSST}}$ (55.82 at 44.1\,kHz and 59.98 at 16\,kHz) while demonstrating competitive performance in DeSync (0.44) and IS$_{\text{PANNs}}$ (17.40) compared to latent-based models~\citep{frieren, wang2024v2a, hunyuanfoley, mmaudio}. These results validate that raw-waveform synthesis can match or even exceed the precision and fidelity of latent-space paradigms. Furthermore, on AudioCaps, our model attains the best FD$_{\text{PANNs}}$ (10.63) and IS$_{\text{PANNs}}$ (12.62) reported to date, rivaling dedicated T2A systems.

In summary, our contributions to direct raw-space audio generation are three-fold:
\begin{itemize}[leftmargin=*, itemsep=2pt, topsep=2pt]
\item \textbf{(i) Streamlined Framework:} we introduce \system, a simplified architecture that synthesizes high-fidelity audio directly in the waveform space through \emph{waveform patchify}, $x$-prediction flow matching, and specialized signal preprocessing, effectively eliminating the need for audio tokenizers.
\item \textbf{(ii) Large-scale Data Curation:} we identify that direct waveform modeling is exceptionally sensitive to data quality and scale, and thus develop an automated pipeline to harvest high-quality, large-scale supervision consisting of multi-modal VT2A samples.
\item \textbf{(iii) Empirical Validation:} we achieve highly competitive results on the VGGSound (VT2A) and AudioCaps (T2A) benchmarks, demonstrating that end-to-end waveform generation reaches performance on par with established latent-based methods in acoustic richness, fidelity, and synchronization.
\end{itemize}

\section{Related Work}
\label{sec:related}

\subsection{Latent-Space Audio Generation}

The landscape of latent-space audio generation is characterized by two main paradigms: \emph{continuous latent modeling} and \emph{discrete codec-based synthesis}. Models such as AudioLDM~\citep{liu2023audioldm, audioldm2}, TANGO~\citep{tango}, and MMAudio~\citep{mmaudio} operate on continuous manifolds learned by audio VAEs~\citep{evans2024fasttimingconditionedlatentaudio}. These frameworks prioritize spectral reconstruction and often incorporate adversarial discriminators from vocoders like HiFi-GAN~\citep{kong2020hifi} or BigVGAN~\citep{lee2023bigvgan} to refine the decoded waveforms. Conversely, systems like AudioGen~\citep{audiogen} and V-AURA~\citep{vaura} leverage discrete neural audio codecs~\citep{defossez2022high,kumar2024high}, where generative modeling is performed over quantized tokens.

While this paradigm bypasses the high-dimensionality of raw audio, it imposes a rigid performance ceiling: the output quality is strictly upper-bounded by the reconstruction fidelity of the pretrained backbone. Critical details, such as high-frequency transients and fine-grained phase information, are often compromised during latent bottlenecking and remain irrecoverable through post-processing. This inherent lossiness motivates the exploration of modeling the audio distribution directly in its native, uncompressed space.

\subsection{Raw-Space Generative Modeling}

Before the dominance of latent-space paradigms, raw waveform modeling was explored through autoregressive and diffusion-based approaches such as WaveNet~\citep{wavenet}, WaveRNN~\citep{wavernn}, WaveGrad~\citep{wavegrad}, and DiffWave~\citep{diffwave}. These methods prove high-fidelity synthesis is feasible without intermediate compression, yet they primarily function as neural vocoders reconstructing waveforms from local spectral features. Consequently, the lack of a mechanism to map global semantic cues directly to raw waveforms limits their use in large-scale multimodal generation.

In the image domain, while early CNNs relied on specialized noise schedules~\citep{chen2023importancenoiseschedulingdiffusion, hoogeboom2023simple}, Transformers often suffer from catastrophic degradation in high-dimensional raw space~\citep{li2026basicsdenoise}. To mitigate this, frameworks like SiD2~\citep{hoogeboom2025simplerdiffusionsid215}, PixelFlow~\citep{chen2025pixelflow}, and PixNerd~\citep{wang2025pixnerdpixelneuralfield} resort to hierarchical designs or specialized heads. Most recently, JiT~\citep{li2026basicsdenoise} succeeds by revisiting the manifold hypothesis~\citep{chapelle2006,JMLR:v11:vincent10a}: since clean data lies on a low-dimensional manifold while noise or velocity spans the entire high-dimensional space, $x$-prediction is fundamentally easier to learn than noise or $v$-prediction. This allows the network to focus on recovering the low-dimensional data structure rather than modeling full-space noise. These advances in vision suggest that the raw-space paradigm, if properly adapted, can overcome the scalability issues previously encountered in audio modeling.

\subsection{Multimodal DiT for Audio Generation}

Video-to-audio (VT2A) generation requires precise temporal synchronization and semantic consistency, leading recent systems to adopt Multimodal Diffusion Transformers (MMDiT)~\citep{sd3} for joint modeling of audio, video, and text. The evolution of these architectures reflects a progression from efficient latent-space synthesis in Frieren~\citep{frieren} to the unified joint-attention paradigm of MMAudio~\citep{mmaudio}, which significantly improved cross-modal alignment. More recently, industrial-scale models~\citep{hunyuanfoley,klingfoley} have pushed performance limits by scaling architectures to dozens of layers and training on massive datasets, such as the 100k hours of video-text-audio samples used in HunyuanVideo-Foley, while utilizing universal latent codecs and enhanced visual modules for high-fidelity synthesis.

Despite these advancements, existing systems remain confined to the compressed latent space. Our work overcomes this by adopting an MMDiT-based architecture that eliminates the latent stage entirely, enabling high-fidelity synthesis directly on raw waveforms.

\section{Method}

The architecture of \system is built on a MultiModal Diffusion Transformer (MMDiT)~\citep{sd3, mmaudio} backbone. Given a multimodal conditioning signal $c$ (video and text), the model employs conditional flow matching to generate the raw waveform $x \in \mathbb{R}^{T}$ directly in observation space. To manage the challenges of high-dimensional audio, we apply \emph{waveform patchify} to reshape the signal for transformer processing and adopt an \emph{$x$-prediction} strategy to ensure stable training.

\subsection{Flow Matching in Waveform Space}
\label{sec:flowmatching}

\paragraph{Conditional Flow Matching.}
We formulate waveform generation using conditional flow matching~\citep{lipman2023flow,liu2023flow,albergo2023building}. Let $x_0 \sim \mathcal{N}(0, I)$ denote Gaussian noise and $x_1$ denote a clean waveform. A continuous interpolation between noise and data is defined as:
\begin{equation}
x_t = (1 - t)x_0 + tx_1, \quad t \in [0, 1],
\end{equation}
with the corresponding target velocity $v^*(x_t, t) = x_1 - x_0$. The goal is to learn a velocity field $v_\theta(x_t, t, c)$ that transports noise to data along this path. While latent-space methods model these flows in a compressed representation, we perform this mapping directly in the waveform space, solving the ODE $\frac{d x_t}{d t} = v_\theta(x_t, t, c)$ during inference to obtain the final waveform.

\paragraph{Prediction Parameterization and Loss.}
We adopt $x$-prediction~\citep{li2026basicsdenoise,salimans2022progressivedistillation}, where network predicts the clean signal:
\begin{equation}
\hat{x}_1 = f_\theta(x_t, t, c).
\end{equation}
The velocity is recovered as $v_\theta = (\hat{x}_1 - x_t) / (1 - t)$. Our default configuration optimizes this $x$-prediction through a $v$-loss:
\begin{equation}
\mathcal{L} = \mathbb{E}_{x_0, x_1, t} \left[ \left\| \frac{\hat{x}_1 - x_t}{1 - t} - \frac{x_1 - x_t}{1 - t} \right\|_2^2 \right].
\end{equation}
This combination ensures that while the network focuses on recovering the data manifold~\citep{chapelle2006}, the objective remains anchored to the flow-matching velocity field. We validate this design choice through ablation experiments in Section~\ref{para:pred}.

\subsection{Model Architecture}
\label{sec:model}

\paragraph{Audio Preprocessing.}
Raw waveforms typically exhibit a sharp, zero-centered distribution with low energy (average RMS often below $0.2$), making them easily masked by noise during training. To mitigate this, we apply \emph{amplitude lifting} by combining RMS normalization and global scaling. Specifically, after converting audio $x \in \mathbb{R}^{T}$ to mono, the lifted waveform $x_{lift}$ is computed as:
\begin{equation}
x_{lift} = s_a \cdot \text{clamp}\left( \frac{r_\star}{rms(x)}x, -1, 1 \right),
\end{equation}
where we empirically set $r_\star = 0.33$ and $s_a = 3.0$ to align the signal scale with the Gaussian noise prior. During inference, the output is rescaled by $1/s_a$ and normalized to $-23$\,LUFS~\citep{ebu2020r128} to ensure perceptually comfortable playback. A visualization of this shift is provided in Appendix~\ref{app:amplitude}.

\begin{figure}[!t]
\centering
\maybeincludegraphics[width=0.9\linewidth]{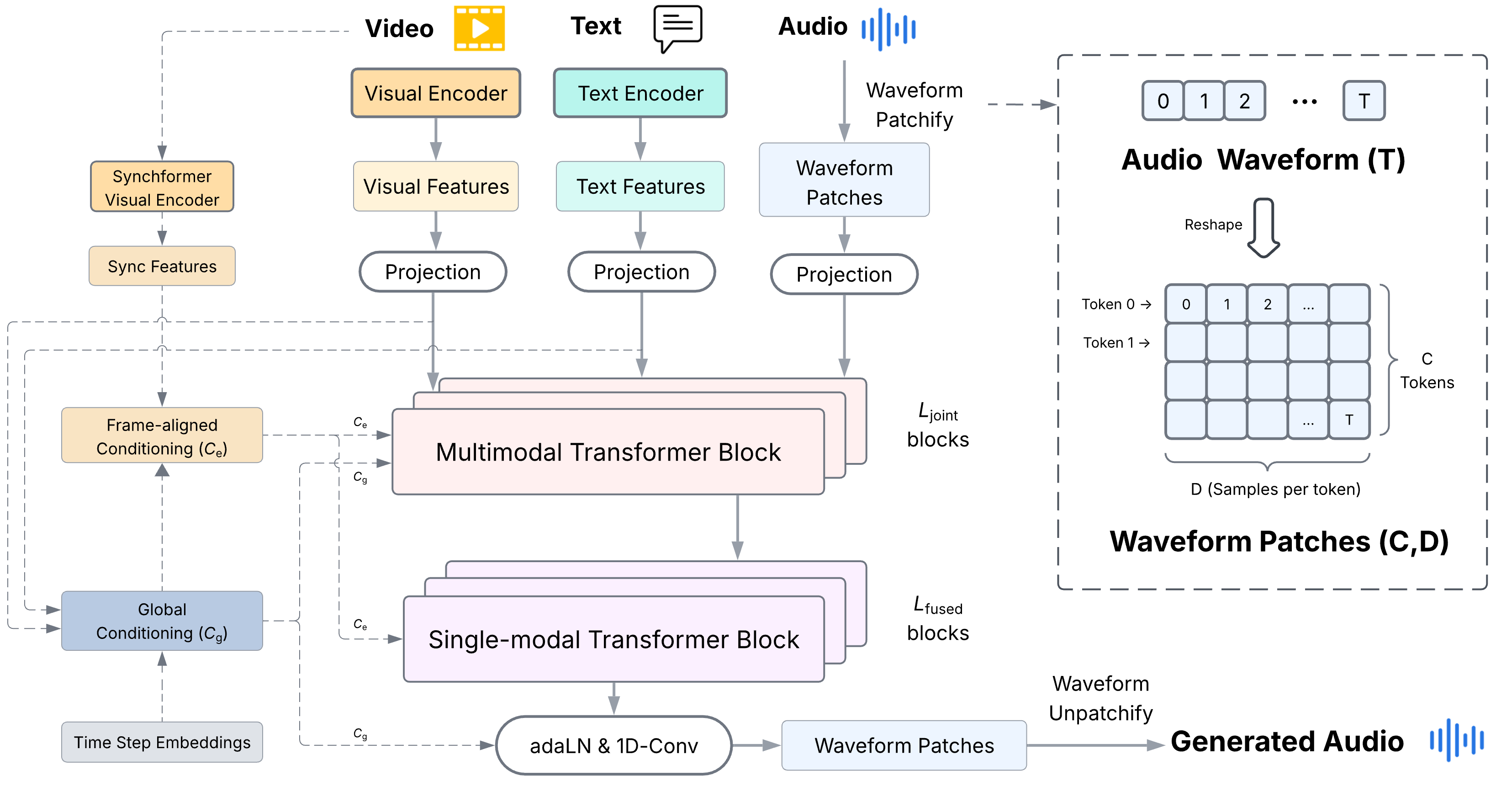}
\vspace{-0.1in}
\caption{\textbf{The \system Architecture.} Raw audio is represented as a 2D patch grid and processed through a series of joint and fused transformer blocks. The model leverages multimodal conditioning ($\mathbf{c}_g$ and $\mathbf{c}_e$) for precise semantic and temporal control during the flow-matching process.}
\label{fig:mmdit}
\end{figure}

\paragraph{Waveform Patchify.}
\label{para:method_patchify}
After preprocessing, raw audio is reshaped into a $C \times D$ grid via \emph{waveform patchify}~(Figure~\ref{fig:mmdit}), where each row serves as a token analogous to the image patchify in ViT~\citep{vit}. The patch dimension $D$ represents the samples per token, defining its temporal granularity. This involves a fundamental trade-off: while smaller $D$ eases the learning of intricate acoustic details, it increases computational complexity ($O(C^2)$); conversely, larger $D$ improves efficiency but increases per-token information density. Ablation studies (Section~\ref{para:patchify}) reveal that increasing the data scale effectively compensates for this modeling difficulty, allowing the network to extract sufficient information even from wider patches.

Our investigations identify $D{=}200$ as the saturation point where performance stabilizes. At $16$\,kHz, this yields $C{=}640$ tokens for an $8$\,s clip, resulting in a $12.5$\,ms granularity---well below the $\sim$25\,ms human auditory resolution threshold~\citep{petrini2009multisensory}. To maintain architectural consistency, we keep $D{=}200$ for $44.1$\,kHz signals ($C{=}1,764$), resulting in an even finer temporal granularity that further reinforces the model's high-fidelity representation. After generation, the grid is reshaped back to a 1D waveform (\emph{waveform unpatchify}). This process is entirely parameter-free and lossless, requiring no learned decoders or neural vocoders.

\paragraph{Multimodal DiT Architecture.}
As shown in Figure~\ref{fig:mmdit}, we adopt the Multimodal Diffusion Transformer (MMDiT)~\citep{sd3,mmaudio} as our backbone, consisting of $L_\mathrm{joint}$ joint blocks for multimodal fusion followed by $L_\mathrm{fused}$ audio-only blocks for waveform refinement. Three input streams enter the joint attention sequence: audio waveform tokens, visual features from a frozen CLIP~\citep{clip} encoder, and text embeddings from a CLIP text encoder. Audio waveform tokens and visual CLIP features are projected into a shared hidden dimension $d$ via convolutional input blocks, whereas text features use a linear projection.

The model employs dual-level conditioning to capture semantic (``what'') and temporal (``when'') cues. A global condition $\mathbf{c}_g$ is formed by summing mean-pooled visual and text features with the flow-matching timestep embedding, providing semantic guidance. To capture precise temporal cues, a frozen Synchformer~\citep{synchformer} extracts synchronization features from video, augmented with learnable per-segment positional embeddings. A frame-aligned condition $\mathbf{c}_e \in \mathbb{R}^{C \times d}$ is obtained by adding $\mathbf{c}_g$ to these synchronization features (upsampled to length $C$ via nearest interpolation), providing frame-level alignment. These conditions are injected into the transformer blocks through AdaLN modulation~\citep{dit}, ensuring robust audio-visual correlation and semantic grounding in the raw waveform space.

Following the transformer blocks, a final output block projects the features from $d$ back to $D$ samples per token via AdaLN and a 1D convolution (kernel size 7). The resulting grid is then reconstructed into a 1D waveform via \emph{waveform unpatchify}. We instantiate two variants based on scale: \textbf{WavFlow-M} ($L_\mathrm{joint}{=}4, L_\mathrm{fused}{=}8, \sim$624M parameters) and \textbf{WavFlow-L} ($L_\mathrm{joint}{=}7, L_\mathrm{fused}{=}14, \sim$1.03B parameters), both sharing hidden dimension $d{=}896$ and 14 attention heads.

\paragraph{Positional Encoding.}
We apply RoPE~\citep{rope} on the queries and keys to inject relative
position information into joint attention; text tokens are excluded
since captions encode unordered semantics rather than temporal
structure. Because the audio and visual CLIP streams run at different
frame rates, applying identical base frequencies would map equivalent
moments in the two streams to mismatched rotary angles. We therefore
multiply the visual stream's RoPE base frequency by the audio-to-visual
ratio $C / N_{\text{clip}}$ (e.g., $10\times$ for $C{=}640,
N_{\text{clip}}{=}64$), so that tokens at the same relative temporal
position receive matching rotary phases.

\subsection{Training and Inference}

\paragraph{Classifier-Free Guidance.}
During training, we independently replace the visual conditioning (visual CLIP and Synchformer features jointly) and text features with learned null embeddings with a 10\% probability. This strategy not only enables classifier-free guidance during inference but also allows \system to support both video-to-audio (VT2A) and text-to-audio (T2A) tasks within a single model. For T2A generation, we simply zero out the visual pathways using the learned null embeddings, reducing the conditioning signal to text alone without any architectural modification.

\paragraph{Inference.}
Generation begins with Gaussian noise sampled in the waveform token space. We solve the learned ODE using an Euler solver with classifier-free guidance (CFG):
\begin{equation}
  \hat{v}_\theta = (1 + w) v_\theta(x_t, t, c) - w v_\theta(x_t, t, \varnothing),
\end{equation}
where $w$ is the guidance scale and $\varnothing$ denotes the null conditions. After the integration, the generated token grid is converted back to a 1D raw waveform via \emph{waveform unpatchify}, requiring no additional learned decoder.

\section{Experiments}
\label{sec:exp}
\subsection{Dataset}
\label{sec:data}

Training generative models directly in raw waveform space imposes significant demands on data scale and quality. In latent-space methods, a pretrained audio encoder leverages extensive audio data (e.g., $\sim$20\,K hours in~\citep{evans2024fasttimingconditionedlatentaudio}) to encode rich acoustic priors, effectively mapping complex waveforms into a compressed manifold that simplifies generative learning. By contrast, waveform-space models must learn intricate acoustic patterns and cross-modal dependencies from scratch, necessitating access to large-scale, high-fidelity audio-visual datasets.

Consequently, we curate a large-scale proprietary media dataset via an automated unified pipeline, constructing robust training sets for both tasks.
\begin{figure}[!t]
\centering
\maybeincludegraphics[width=1.0\linewidth]{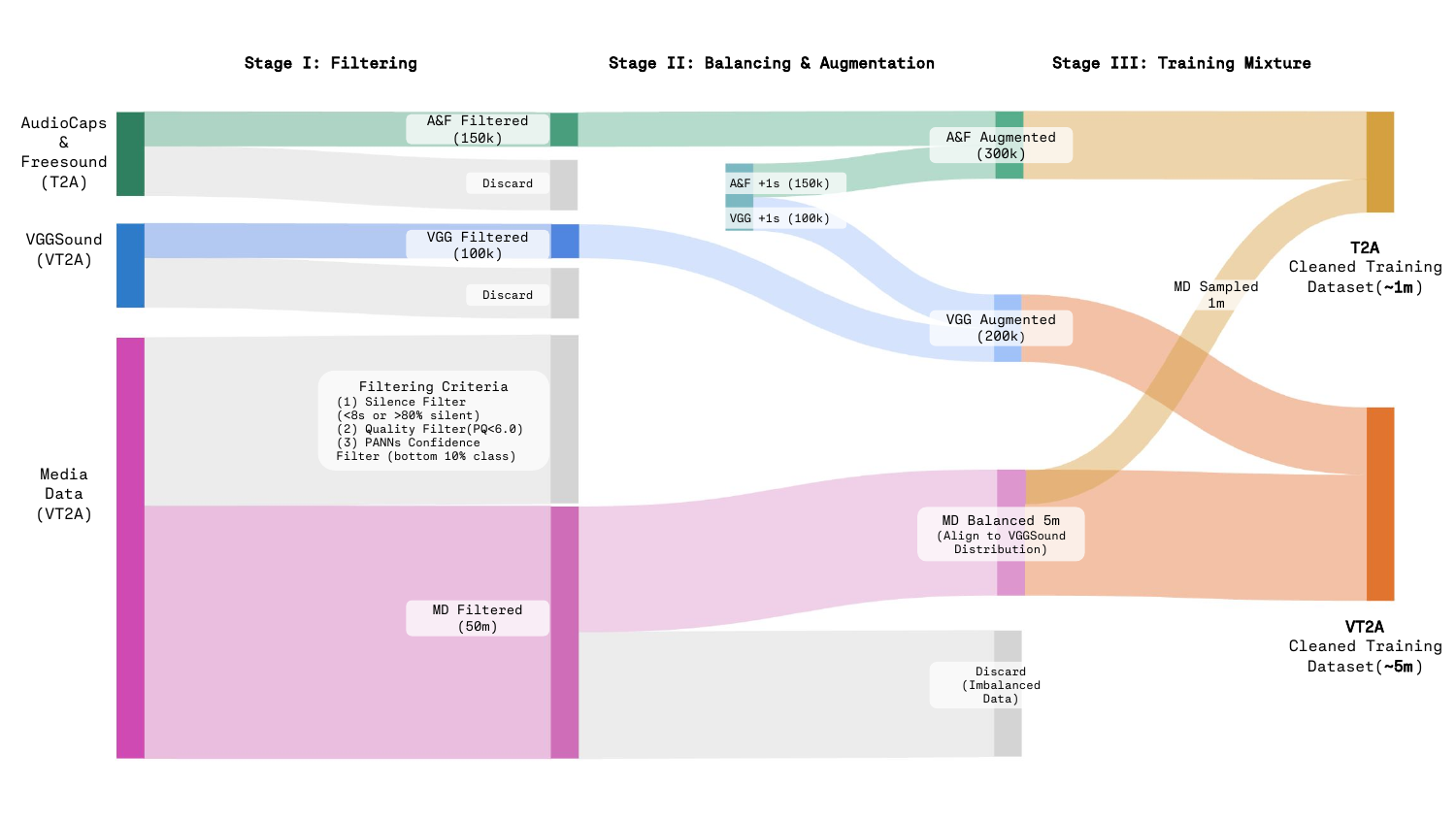}
\vspace{-0.1in}
\caption{Overview of multi-stage data curation from raw datasets to final high quality, balanced training mixtures.}
\label{fig:datamix}
\end{figure}

\paragraph{Data Curation Pipeline.}
As illustrated in Figure~\ref{fig:datamix}, our pipeline unifies VT2A and T2A samples through three stages. For open-source data, we utilize VGGSound alongside AudioCaps~\citep{kim2019audiocaps} and Freesound~\citep{font2013freesound}. Initially, we apply multi-stage filtering across all sources: extracting $8$\,s segments and discarding samples with $>80$\% silence, low aesthetic scores (PQ $<6.0$ via audiobox-aesthetics~\citep{tjandra2025aes}), or low classification confidence (bottom $10$\% via PANNs~\citep{panns}). This process yields roughly $50$\,M filtered media clips, $100$\,K VGGSound samples, and $150$\,K high-quality T2A samples.

Subsequently, we balance and augment the filtered data. The curated media clips are category-aligned with VGGSound to form a balanced pool of $5$\,M samples. For the smaller VGGSound and public T2A sets, we apply temporal augmentation by extracting two overlapping $8$\,s chunks starting at $0$\,s and $1$\,s, respectively, to double their size to $200$\,K and $300$\,K. Ultimately, these sources are merged into our final mixtures: a \textbf{VT2A set} combining the $5$\,M balanced media pool with augmented VGGSound, and a \textbf{T2A set} mixing the $300$\,K augmented public T2A samples with $1$\,M clips randomly sampled from the same high-quality media corpus. This ensures a consistent data distribution across both tasks.

\subsection{Experimental Setup}
\label{subsec:setup}

\paragraph{Training.}
Training configurations are detailed in Table~\ref{tab:app_training}. Models are trained on $8$\,s clips using flow-matching with $x$-prediction and $v$-loss, sampling timesteps from a logit-normal distribution~\citep{sd3}. Raw audio is tokenized via \emph{waveform patchify} ($D{=}200$), resulting in sequence lengths of $640$ tokens for $16$\,kHz and $1{,}764$ tokens for $44.1$\,kHz. We instantiate two $16$\,kHz variants (\textbf{WavFlow-M-16kHz}, \textbf{WavFlow-L-16kHz}) and one $44.1$\,kHz model (\textbf{WavFlow-L-44.1kHz}).

The $16$\,kHz models are trained from scratch for $400$\,epochs, a convergence point identified by monitoring validation metrics (see Figure~\ref{fig:convergence}). Our primary VT2A model utilizes the $\sim$5M mixture with a global batch size of $10{,}752$, while the T2A model uses the $\sim$1M mixture with a batch size of $8{,}192$. For high-fidelity synthesis, the $44.1$\,kHz Large model is fine-tuned (SFT)~\citep{yosinski2014transferable} from the converged $16$\,kHz checkpoint. All stages use the AdamW optimizer with a constant learning rate of $1\times10^{-4}$ ($1\times10^{-5}$ for SFT), a $20$-epoch linear warmup, and an EMA decay of $0.9999$.

\paragraph{Evaluation Metrics.}
We evaluate VT2A on the VGGSound test set ($15$\,K videos) and T2A on AudioCaps ($4.8$\,K samples). We run inference with an ODE solver using $50$\,steps and a CFG scale of $4.5$ (see Appendix~\ref{app:inference}). We assess performance across three dimensions:
\begin{itemize}[leftmargin=*, itemsep=2pt, topsep=2pt]
\item \textbf{(i) Acoustic Quality:} We measure Fr\'echet Distance (FD), KL divergence, and Inception Score (IS) using features from PANNs~\citep{panns} and PaSST~\citep{koutini22passt} classifiers.
\item \textbf{(ii) Semantic Alignment:} The semantic alignment is evaluated via ImageBind (IB)~\citep{girdhar2023imagebind} for audio-visual correspondence and CLAP~\citep{laionclap2023} for text-audio consistency.
\item \textbf{(iii) Temporal Synchronization:} Timing accuracy between visual events and audio onsets is quantified using the DeSync metric~\citep{synchformer}.
\end{itemize}

\subsection{Comparison with State-of-the-arts}
\label{sec:mainexp}

\paragraph{Video-to-Audio Generation.}
Table~\ref{tab:vt2a_results_full} summarizes the results on VGGSound-Test set. Even with a medium-sized backbone at 16\,kHz, \system{} surpasses established latent-based systems---including Frieren~\citep{frieren}, V2A-Mapper~\citep{wang2024v2a}, and HunyuanVideo-Foley~\citep{hunyuanfoley}---across multiple acoustic and synchronization metrics (e.g., FD$_{\text{PANNs}}$: 6.37, IS$_{\text{PANNs}}$: 17.24, and DeSync: 0.47), closely approaching the state-of-the-art MMAudio-L-44.1kHz~\citep{mmaudio} baseline despite the absence of a pretrained neural codec.

Scaling to \system{}-L-16kHz yields consistent improvements, surpassing MMAudio-L-44.1kHz in distributional fidelity (FD$_{\text{PaSST}}$: 59.98 vs.\ 60.60) while matching its performance in perceptual and alignment metrics (IS$_{\text{PANNs}}$: 17.40, DeSync: 0.44). This validates that raw-waveform modeling can attain the same level of temporal precision and acoustic quality as sophisticated latent-space methods. Furthermore, \system{}-L-44.1kHz, fine-tuned from the 16\,kHz Large checkpoint, pushes distributional fidelity even further, achieving the best-reported FD$_{\text{PaSST}}$ (55.82) across all compared methods while maintaining high synchronization (DeSync: 0.46). Collectively, these results confirm that with high-quality, large-scale data, end-to-end raw waveform synthesis can eliminate the need for intermediate latent representations without sacrificing performance.

\begin{table*}[!h]
\caption{Comparison with state-of-the-art methods~\citep{frieren, wang2024v2a, hunyuanfoley,mmaudio} on VGGSound-Test for video-text-to-audio (VT2A) generation. Lower is better for FD, KL and DeSync; higher is better for IS, IB and CLAP. The best and second-best results are highlighted in \textbf{bold} and \underline{underlined}, respectively.}
\centering
\small
\label{tab:vt2a_results_full}
\renewcommand{\arraystretch}{1.1}
\setlength{\tabcolsep}{3pt}
\begin{tabular}{lcccccccc}
\toprule
Method
& FD$_{\text{PANNs}}\downarrow$
& FD$_{\text{PaSST}}\downarrow$
& KL$_{\text{PANNs}}\downarrow$
& IS$_{\text{PANNs}}\uparrow$
& IB$\uparrow$
& DeSync$\downarrow$
& CLAP$\uparrow$
& Params \\
\midrule
Frieren\textsuperscript{$\dagger$}
& 11.45 & 106.10 & 2.73 & 12.25 & 0.23 & 0.85 & 0.11 & 159M \\

V2A-Mapper\textsuperscript{$\dagger$}
& 8.40 & 84.57 & 2.69 & 12.47 & 0.23 & 1.23 & 0.11 & 229M \\

HunyuanVideo-Foley\textsuperscript{$\ast$}
& 10.53 & 97.85 & 2.02 & 14.99 & \underline{0.32} & 0.54 & \textbf{0.23} & -- \\

MMAudio-L-44.1kHz\textsuperscript{$\dagger$}
& \textbf{4.72} & 60.60 & \textbf{1.65} & \textbf{17.40} & \textbf{0.33} & \textbf{0.44} & \underline{0.22} & 1.03B \\

\midrule
WavFlow-M-16kHz
& 6.37 & 62.64 & 1.68 & \underline{17.24} & 0.30 & 0.47 & 0.21 & 624M \\

WavFlow-L-16kHz
& 5.86 & \underline{59.98} & \underline{1.66} & \textbf{17.40} & 0.31 & \textbf{0.44} & \underline{0.22} & 1.03B \\

WavFlow-L-44.1kHz
& \underline{5.25} & \textbf{55.82} & 1.73 & 15.05 & 0.31 & \underline{0.46} & 0.19 & 1.03B \\
\bottomrule
\end{tabular}
\par\noindent{\footnotesize All methods are evaluated on the same VGGSound test split from the MMAudio~\citep{mmaudio} benchmark, utilizing original videos and native class labels as captions to ensure a fair comparison. Due to the difference in semantic granularity (sparse labels vs. dense captions), we exclude direct comparisons with models relying on LLM-refined captions. \textsuperscript{$\dagger$}: results taken from the MMAudio paper. \textsuperscript{$\ast$}: reproduced by using their open-source checkpoints on the same test set.}
\end{table*}

\begin{table}[!h]
\caption{Comparison with state-of-the-art methods~\citep{audioldm2,tango,tango2,makeanaudio,makeanaudio2,genau,mmaudio} on AudioCaps-Test for text-to-audio (T2A) generation.}
\centering
\small
\label{tab:t2a_results}
\vspace{-5pt}
\renewcommand{\arraystretch}{1.1}
\setlength{\tabcolsep}{4pt}
\begin{tabular}{lccccc}
\toprule
Method & Params & FD$_{\text{PANNs}}\downarrow$ & FD$_{\text{VGG}}\downarrow$ & IS$_{\text{PANNs}}\uparrow$ & CLAP$\uparrow$ \\
\midrule
AudioLDM 2-L & 712M & 32.50 & 5.11 & 8.54 & 0.21 \\
TANGO & 866M & 26.13 & 1.87 & 8.23 & 0.19 \\
TANGO 2 & 866M & 19.77 & 2.74 & 8.45 & 0.26 \\
Make-An-Audio & 453M & 27.93 & 2.59 & 7.44 & 0.21 \\
Make-An-Audio 2 & 937M & 15.34 & \underline{1.27} & 9.58 & 0.25 \\
GenAU-Large & 1.25B & 16.51 & \textbf{1.21} & 11.75 & \underline{0.29} \\
MMAudio-L-44.1kHz & 1.03B & \underline{15.04} & 4.03 & \underline{12.08} & \textbf{0.35} \\
\midrule
WavFlow-M-16kHz & 624M & \textbf{10.63} & 1.58 & \textbf{12.62} & 0.24 \\
\bottomrule
\end{tabular}
\vspace{0.05in}
\par\noindent{\footnotesize Baseline results are quoted from~\citep{mmaudio} for a fair comparison, as we adopt the same test splits.}

\end{table}
\paragraph{Text-to-Audio Generation.}
To further validate the versatility of our VAE-free approach, we evaluate WavFlow-M-16kHz on the AudioCaps text-to-audio benchmark. As shown in Table~\ref{tab:t2a_results}, despite being a unified model rather than one specialized for T2A, our system achieves competitive acoustic quality across all compared methods. Specifically, it attains the lowest FD$_{\text{PANNs}}$ (\textbf{10.63}) and the highest IS$_{\text{PANNs}}$ (\textbf{12.62}), outperforming dedicated latent-space models and the previous state-of-the-art MMAudio. These results demonstrate that the intricate acoustic patterns learned directly in raw waveform space generalize effectively across different input modalities. More importantly, this cross-task success reinforces our primary conclusion: high-fidelity synthesis can be achieved without intermediate latent representations or task-specific vocoders.

\subsection{Ablation Studies}
Unless specified, all ablations are conducted using WavFlow-M-16k trained on a mixture of 1M media data and VGGSound, following the default configurations in Appendix~\ref{app:training}. All reported results are evaluated on the VGGSound-Test set.

\begin{figure}[!h]
\centering
\maybeincludegraphics[width=1.0\linewidth]{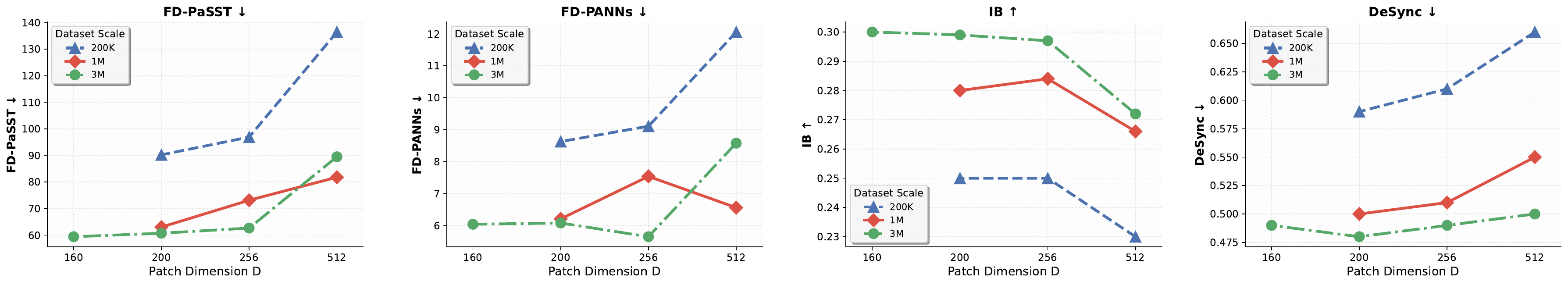}
\vspace{-0.2in}
\caption{\textbf{Patchify granularity vs.\ data scale across key metrics.} Note the performance degradation at $D{=}512$ when scaling to 3M, indicating a capacity bottleneck for low-granularity tokens.}
\label{fig:tokenization}
\end{figure}

\paragraph{Patchify Granularity Analysis.}
\label{para:patchify}
Figure~\ref{fig:tokenization} and Table~\ref{tab:tokenization} reveal an interplay between the patch dimension $D$ and data scale. Reducing $D$ refines the temporal resolution of raw waveform tokens by increasing the token count $C$. In the low-data regime (200K VGGSound), this finer granularity is the most effective lever for quality: shrinking $D$ from 512 to 200 drives a dramatic improvement in FD$_{\text{PaSST}}$ (136.45 $\to$ 90.24) and DeSync (0.66 $\to$ 0.59). This suggests that high-resolution tokenization is essential for capturing intricate acoustic structures when training samples are sparse.

Conversely, increasing data scale can partially compensate for the modeling difficulty of a larger $D$. For a fixed $D{=}512$, expanding the dataset from 200K to 1M significantly improves FD$_{\text{PaSST}}$ (136.45 $\to$ 81.81). However, at 3M samples, this coarse configuration suffers from a \emph{capacity bottleneck}, where performance degrades to 89.51, indicating that large $D$ values eventually fail to represent the complex information inherent in larger datasets. In contrast, with sufficient granularity ($D \le 256$), increasing data scale consistently improves results. Notably, at 3M samples, the performance gain from further reducing $D$ from 200 to 160 is marginal (60.75 vs. 59.43), identifying $D{=}200$ as the \emph{saturation point} where computational efficiency and synthesis quality are optimally balanced. Based on these findings, we adopt $D{=}200$ ($C{=}640$) as our default configuration.

\vspace{-0.15in}
\begin{table}[H]
\caption{\textbf{Impact of patchify granularity ($C \times D$) on waveform synthesis across varying data scales.} Results demonstrate the trade-off between temporal resolution and data diversity.}
\centering
\small
\label{tab:tokenization}
\vspace{-4pt}
\renewcommand{\arraystretch}{1.1}
\setlength{\tabcolsep}{3.5pt}
\begin{tabular}{llcccccc}
\toprule
Data Scale & $C \times D$
& FD$_{\text{PaSST}}\downarrow$
& FD$_{\text{PANNs}}\downarrow$
& KL$_{\text{PANNs}}\downarrow$
& IS$_{\text{PANNs}}\uparrow$
& IB$\uparrow$
& DeSync$\downarrow$ \\
\midrule
\multirow{3}{*}{\shortstack[l]{200K\\(VGGSound)}}
& 250 $\times$ 512 & 136.45 & 12.06 & 1.88 & 10.86 & 0.23 & 0.66 \\
& 500 $\times$ 256 &  96.91 &  9.11 & 1.84 & 12.56 & 0.25 & 0.61 \\
& 640 $\times$ 200 &  90.24 &  8.63 & 1.83 & 13.45 & 0.25 & 0.59 \\
\midrule
\multirow{3}{*}{1M}
& 250 $\times$ 512 &  81.81 &  6.56 & 1.78 & 13.66 & 0.27 & 0.55 \\
& 500 $\times$ 256 &  73.16 &  7.54 & 1.91 & 16.18 & 0.28 & 0.51 \\
& 640 $\times$ 200 &  63.05 &  6.21 & 1.76 & 15.58 & 0.28 & 0.50 \\
\midrule
\multirow{4}{*}{3M}
& 250 $\times$ 512 &  89.51 &  8.58 & 1.96 & 15.60 & 0.27 & 0.50 \\
& 500 $\times$ 256 &  62.69 &  5.65 & 1.68 & 15.80 & 0.30 & 0.49 \\
& 640 $\times$ 200 &  60.75 &  6.08 & 1.74 & 16.56 & 0.30 & 0.48 \\
& 800 $\times$ 160 &  59.43 &  6.04 & 1.73 & 16.43 & 0.30 & 0.49 \\
\bottomrule
\end{tabular}
\vspace{-0.05in}
\end{table}
\vspace{-0.1in}

\paragraph{Prediction Target and Loss Formulation.}
\label{para:pred}
Table~\ref{tab:ablation_pred} compares prediction targets ($x$ vs.\ $v$) and loss formulations. The results show that $x$-prediction consistently outperforms $v$-prediction across all metrics, likely because it provides a more structured objective that better conforms to the manifold hypothesis~\citep{chapelle2006, li2026basicsdenoise}. It is worth noting that both targets remain viable for direct raw-waveform generation within our framework. This universal effectiveness can be attributed to the low per-token dimensionality ($D{=}200$) relative to the model's hidden dimension ($d{=}896$), which provides sufficient capacity to represent either target without a latent bottleneck.

Among the $x$-prediction variants, $x$-pred with $v$-loss achieves the best FD$_{\text{PaSST}}$ (63.05) and IS$_{\text{PANNs}}$ (15.58), whereas $x$-pred with $x$-loss shows slight advantages in FD$_{\text{PANNs}}$ (4.86). Given that $v$-loss superiorly balances generative diversity (IS$_{\text{PANNs}}$) and high-frequency fidelity---as reflected by FD$_{\text{PaSST}}$, which leverages the 32\,kHz-aware PaSST to capture a wider frequency range than PANNs (operates on 16\,kHz) ---we adopt $x$-prediction with $v$-loss as our default.

\begin{table}[H]
\caption{\textbf{Ablation on WavFlow's Flow Matching objectives.} Results indicate that $x$-prediction with $v$-loss provides the optimal balance between high-level feature similarity (FD$_{\text{PaSST}}$) and generative diversity (IS$_{\text{PANNs}}$).}
\centering
\small
\label{tab:ablation_pred}
\vspace{-4pt}
\renewcommand{\arraystretch}{1.1}
\setlength{\tabcolsep}{3pt}
\begin{tabular}{lcccccc}
\toprule
Setting
& FD$_{\text{PaSST}}\downarrow$
& FD$_{\text{PANNs}}\downarrow$
& KL$_{\text{PANNs}}\downarrow$
& IS$_{\text{PANNs}}\uparrow$
& IB$\uparrow$
& DeSync$\downarrow$\\
\midrule
$v$-pred + $v$-loss
& 77.19 & 6.38 & 1.75 & 13.48 & 0.27 & 0.53 \\
$x$-pred + $x$-loss
& 72.70 & 4.86 & 1.72 & 13.99 & 0.29 & 0.50 \\
$x$-pred + $v$-loss
& 63.05 & 6.21 & 1.76 & 15.58 & 0.28 & 0.50 \\
\bottomrule
\end{tabular}
\vspace{-0.05in}
\end{table}

\paragraph{Raw-waveform Preprocessing.}
Table~\ref{tab:ablation_preprocess} validates our preprocessing pipeline, showing that both RMS normalization and amplitude scaling are essential for signal quality. Without scaling (1.0$\times$), omitting RMS normalization severely degrades FD$_{\text{PaSST}}$ (65.83 $\to$ 81.26) and DeSync (0.49 $\to$ 0.57). At 3.0$\times$ scale, the enhanced signal strength naturally narrows the performance gap, yet the combination of both techniques still yields the best overall results in diversity and fidelity (e.g., IS$_{\text{PANNs}}$ of 15.58). We hypothesize that proper scaling helps the raw signal better align with the Gaussian prior during flow matching, whereas insufficient scaling (1.0$\times$) leads to suboptimal optimization. Consequently, we adopt RMS normalization with a 3.0$\times$ scale as our default.

\begin{table}[!h]
\caption{\textbf{Ablation of raw-waveform preprocessing.} Results indicate that RMS normalization with 3.0$\times$ scaling yields optimal performance.}
\centering
\small
\label{tab:ablation_preprocess}
\vspace{-4pt}
\renewcommand{\arraystretch}{1.1}
\setlength{\tabcolsep}{4pt}
\begin{tabular}{llcccccc}
\toprule
Category & Setting
& FD$_{\text{PaSST}}$$\downarrow$
& FD$_{\text{PANNs}}$$\downarrow$
& KL$_{\text{PANNs}}\downarrow$
& IS$_{\text{PANNs}}\uparrow$
& IB$\uparrow$
& DeSync$\downarrow$ \\
\midrule
\multirow{2}{*}{\shortstack[l]{RMS Norm.\\(at 1.0$\times$ scale)}}
& w/    & 65.83 & 6.03 & 1.73 & 13.32 & 0.28 & 0.49 \\
& w/o & 81.26 & 8.69 & 1.91 & 11.64 & 0.24 & 0.57 \\
\midrule
\multirow{2}{*}{\shortstack[l]{RMS Norm.\\(at 3.0$\times$ scale)}}
& w    & 63.05 & 6.21 & 1.76 & 15.58 & 0.28 & 0.50 \\
& w/o & 64.23 & 6.93 & 1.79 & 13.84 & 0.26 & 0.52 \\
\bottomrule
\end{tabular}
\vspace{-0.05in}
\end{table}

\section{Conclusion and Limitations}
\label{sec:conclusion}

We present \system{}, a flow-matching framework for high-fidelity audio generation directly in raw-waveform space. By combining waveform patchify, $x$-prediction, and specific signal scaling with extensive data, \system{} effectively stabilizes training and models complex raw signals. It achieves highly competitive performance on VT2A and T2A benchmarks, matching or exceeding established latent-based systems.
\paragraph{Limitations.}
\system{} currently lacks explicit speech or singing synthesis, as generated vocalizations do not constitute meaningful language. Extending to these domains requires finer linguistic granularity and larger speech datasets. By incorporating larger-scale corpora and fine-grained linguistic captions, this framework could be extended to jointly model environmental sounds and human speech, offering a robust and efficient alternative for future generative research.

\clearpage
\bibliographystyle{assets/plainnat}
\bibliography{main}

\clearpage
\beginappendix

\section{Training Details}
\label{app:training}

Table~\ref{tab:app_training} summarizes the full training configurations for all \system{} variants. All models are trained on NVIDIA H100 GPUs and share the same optimizer (AdamW with $\beta_1{=}0.9$, $\beta_2{=}0.95$), EMA decay of $0.9999$, gradient clipping at $1.0$, and BF16 mixed precision. In our main experiments, $16$\,kHz VT2A models are trained from scratch with a learning rate of $1\times10^{-4}$ and a global batch size of $10{,}752$ on the mixture described in Sec.~\ref{sec:data} (comprising $5$\,M media data and $200$\,K augmented VGGSound). Within each epoch, every dataset is traversed exactly once to ensure balanced supervision. The T2A model follows the same architecture but is trained separately on the $1$\,M T2A mixture with a batch size of $8{,}192$. WavFlow-L-44.1kHz is obtained via supervised fine-tuning from the converged $16$\,kHz checkpoint on $44.1$\,kHz data, using a reduced learning rate of $1\times10^{-5}$ and a batch size of $1{,}536$.

\begin{table}[H]
\caption{Training hyperparameters for all \system variants. All models use 16\,kHz sample rate except WavFlow-L-44k (44.1\,kHz).}
\centering
\small
\label{tab:app_training}
\vspace{0.05in}
\renewcommand{\arraystretch}{1.1}
\setlength{\tabcolsep}{3pt}
\begin{tabular}{lcccc}
\toprule
 & \multicolumn{3}{c}{VT2A} & T2A \\
\cmidrule(lr){2-4} \cmidrule(lr){5-5}
Hyperparameter & WavFlow-M-16k & WavFlow-L-16k & WavFlow-L-44k & WavFlow-M-16k \\
\midrule
Backbone\textsuperscript{\dag} & MMDiT-M & MMDiT-L & MMDiT-L & MMDiT-M \\
Joint layers $L_\mathrm{joint}$ & 4 & 7 & 7 & 4 \\
Single layers $L_\mathrm{fused}$ & 8 & 14 & 14 & 8 \\
Hidden dim $d$ & 896 & 896 & 896 & 896 \\
Attention heads & 14 & 14 & 14 & 14 \\
Parameters & 624M & 1.03B & 1.03B & 624M \\
\midrule
Training data & \makecell{Media data 5M\\+ VGG 200K} & \makecell{Media data 5M\\+ VGG 200K} & \makecell{VGG 200K\\(44.1\,kHz)} & \makecell{Media data 1M\\+ T2A open source 300K} \\
Optimizer & AdamW & AdamW & AdamW & AdamW \\
$(\beta_1, \beta_2)$ & (0.9, 0.95) & (0.9, 0.95) & (0.9, 0.95) & (0.9, 0.95) \\
Learning rate & 1e-4 & 1e-4 & 1e-5 & 1e-4 \\
LR schedule & Constant & Constant & Constant & Constant \\
Warmup & 20 epochs & 20 epochs & 20 epochs & 20 epochs \\
Global batch size & 10752 & 10752 & 1536 & 8192 \\
Training epochs & 400 & 400 & 650 & 400 \\
Initialization & Scratch & Scratch & SFT from L-16k & Scratch \\
EMA decay & 0.9999 & 0.9999 & 0.9999 & 0.9999 \\
Gradient clipping & 1.0 & 1.0 & 1.0 & 1.0 \\
Precision & BF16 & BF16 & BF16 & BF16 \\
\midrule
Patchify ($C \times D$) & $640 \times 200$ & $640 \times 200$ & $1764 \times 200$ & $640 \times 200$ \\
Audio scale & 3.0 & 3.0 & 3.0 & 3.0 \\
CFG drop rate (video / text) & 10\% / 10\% & 10\% / 10\% & 10\% / 10\% & -- / 10\% \\
ODE steps (inference) & 50 & 50 & 50 & 50 \\
CFG strength (inference) & 4.5 & 4.5 & 4.5 & 4.5 \\
\bottomrule
\end{tabular}
\vspace{0.05in}
\par\noindent{\footnotesize \textsuperscript{\dag}Multimodal DiT backbone adopted from MMAudio~\citep{mmaudio}. ``M'' and ``L'' denote the medium and large configurations.}
\end{table}
\vspace{-0.15in}

\begin{figure}[H]
\centering
\maybeincludegraphics[width=1.0\linewidth]{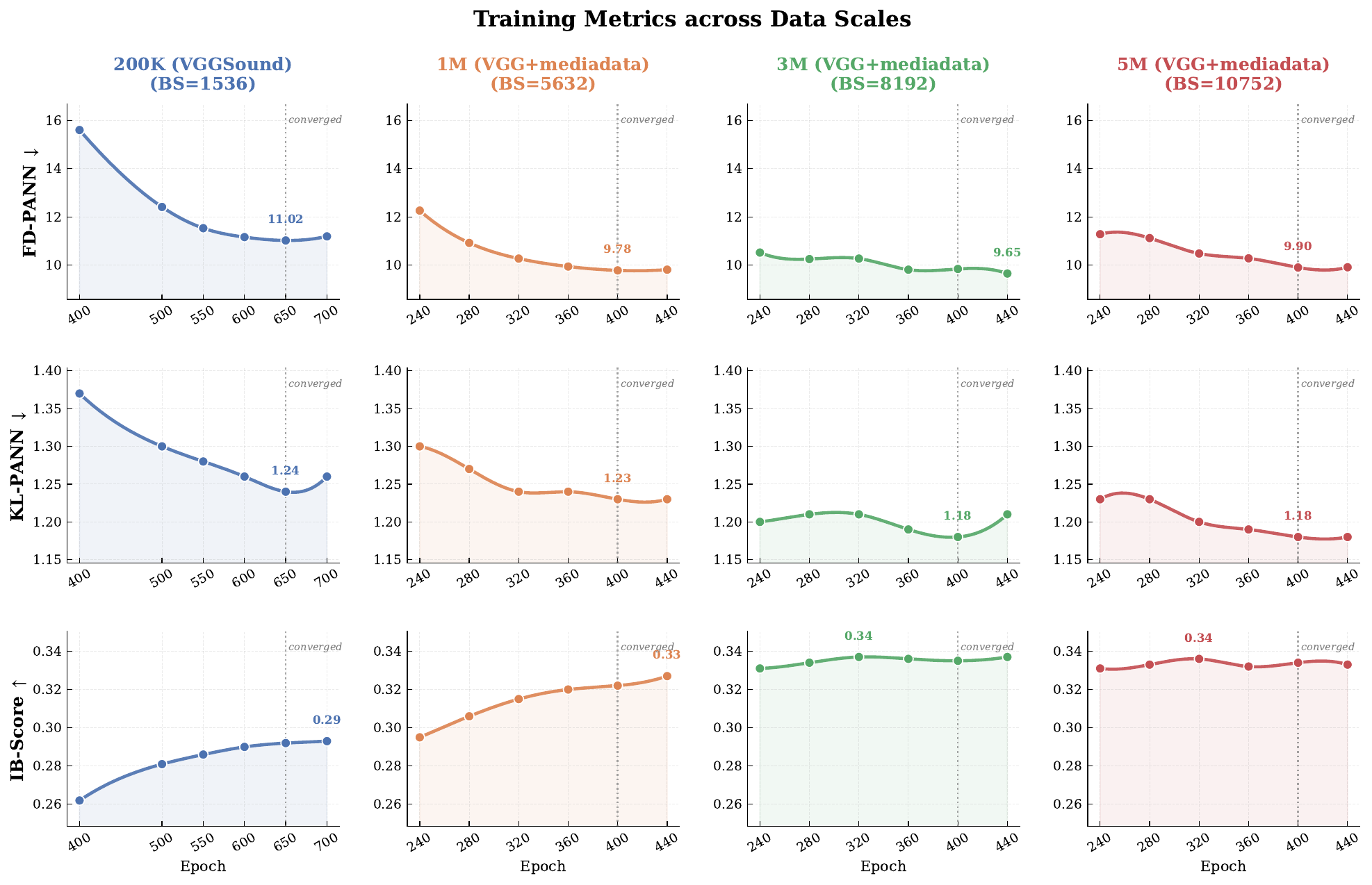}
\vspace{-0.1in}
\caption{\textbf{Metrics on the VGGSound validation set as a function of training epochs across four data scales:} $200$\,K VGGSound (BS=$1{,}536$), $1$\,M (BS=$5{,}632$), $3$\,M (BS=$8{,}192$), and $5$\,M (BS=$10{,}752$). Top row: FD$_{\text{PANNs}}\!\downarrow$; middle row: KL$_{\text{PANNs}}\!\downarrow$; bottom row: IB\,$\uparrow$. Models trained with $\ge 1$\,M samples converge around $400$\,epochs, while the $200$\,K setting requires $\sim$$650$\,epochs due to its smaller batch size. Dashed lines indicate the convergence points adopted for final training. Here, BS denotes the global batch size.}
\label{fig:convergence}
\end{figure}

To determine the appropriate number of training epochs, we budget each convergence study to approximately $84$\,hours and scale the global batch size proportionally with dataset size so that all runs complete within this timeframe. We monitor validation metrics on the VGGSound validation set ($\sim$$2$\,K samples) throughout training across four data scales ($200$\,K, $1$\,M, $3$\,M, and $5$\,M). As shown in Figure~\ref{fig:convergence}, models trained with $\ge 1$\,M data and correspondingly larger batch sizes ($5{,}632$--$10{,}752$) converge by approximately $400$\,epochs, with FD$_{\text{PANNs}}$, KL$_{\text{PANNs}}$, and IB scores all plateauing beyond this point. We therefore adopt $400$\,epochs as the default for both main and ablation experiments at this scale. The $200$\,K VGGSound-only run requires approximately $650$\,epochs to stabilize due to its much smaller batch size of $1{,}536$, and the $44.1$\,kHz fine-tuning setting similarly uses $650$\,epochs to reach convergence under its reduced learning rate. Additionally, the global batch sizes labeled in Figure~\ref{fig:convergence} correspond to the fixed settings utilized in our ablation studies.
\vspace{-0.15in}

\section{Audio Amplitude Distribution}
\label{app:amplitude}

\begin{figure}[!t]
\centering
\includegraphics[width=1.0\linewidth]{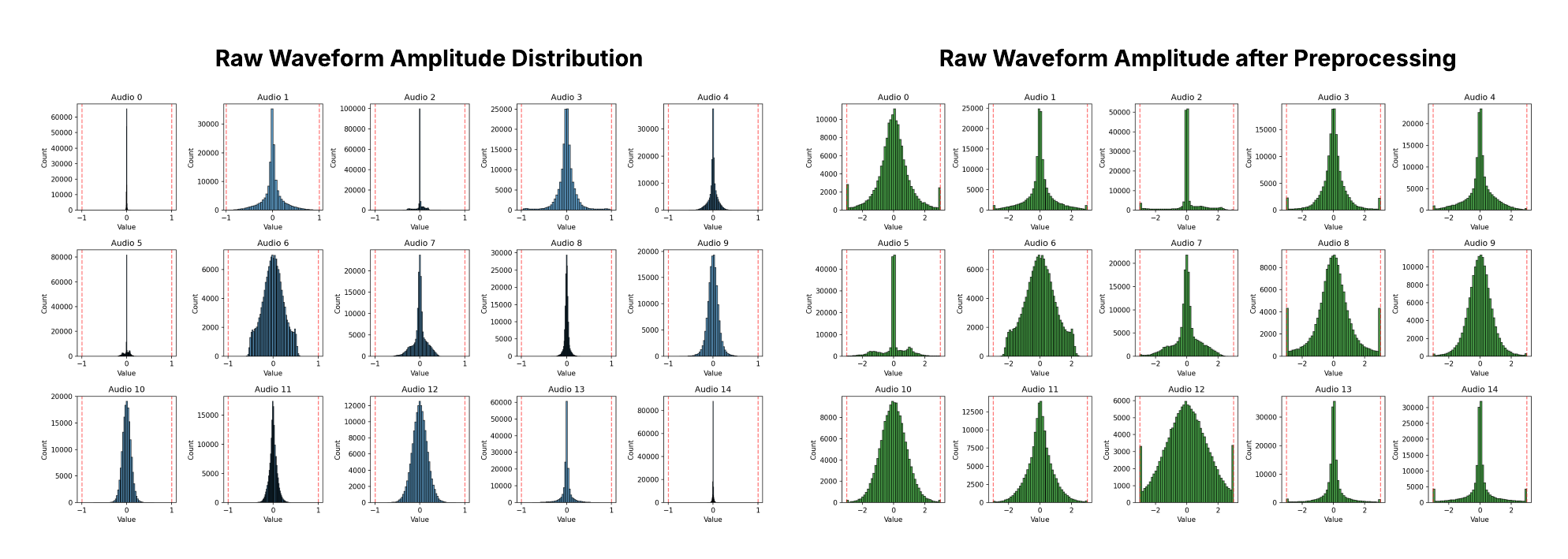}
\caption{\textbf{Amplitude histograms of $15$ randomly sampled audio clips before (left, blue) and after (right, green) preprocessing} (RMS normalization to $0.33$ followed by $\times 3.0$ scaling). Raw waveforms exhibit sharp zero-centered peaks with widely varying energy across clips. After preprocessing, the amplitude values are more consistent, lifting the signal to ensure it is not submerged by the Gaussian noise prior during flow matching.}
\label{fig:app_amplitude}
\end{figure}
\vspace{-0.15in}

Waveform-space amplitudes exhibit a wide dynamic range, yet values are heavily concentrated near zero, with empirical observations indicating that most waveform RMS levels typically remain below $0.2$. The left panel of Figure~\ref{fig:app_amplitude} illustrates this by plotting the amplitude histograms of $15$ randomly sampled clips. Quiet samples, such as Audio~0, 2, and 14, occupy only a minimal fraction of the available dynamic range, while others, such as Audio~4, 9 and 11, show broader yet still sharply zero-centered distributions. These low-energy signals possess negligible magnitudes and are easily submerged by noise during training. This further results in vanishingly small loss values, providing insufficient gradient information and making the denoising objective difficult to optimize.

In contrast, after RMS normalization to a target level of $0.33$ followed by $\times 3.0$ amplitude scaling (right panel), the signals are effectively lifted, with distributions spreading more broadly across the $[-3, 3]$ range. While the resulting distributions remain non-uniform, their amplitude statistics are much better aligned with the standard $N(0, 1)$ noise prior than the raw waveforms. This redistribution ensures that the signal remains discernible even at high noise levels, rendering the denoising objective significantly more tractable and training more stable. The quantitative impact of this preprocessing is validated through ablation experiments (Table~\ref{tab:ablation_preprocess}).

\section{Data Mixture and Rationale for VT2A}
\label{app:datamix}

This section describes our process for selecting the optimal training data mixture for VT2A generation and the reasoning behind our final configuration. We consider three primary data sources: (1) \textbf{VGGSound (200K)}, which consists of paired video-audio samples with \textbf{sparse labels} (e.g., ``dog barking''); (2) \textbf{Open-source T2A data (300K)}, which includes audio-only samples from FreeSound and AudioCaps featuring \textbf{fine-grained captions} (e.g., ``a vehicle engine accelerating then running on idle''); and (3) \textbf{Media data}, a large-scale collection of $5$\,M high-quality proprietary video-audio pairs derived from the MovieGen~\citep{polyak2025moviegencastmedia} training subset, featuring fine-grained captions comparable in detail to the Open-source T2A data. For experimental efficiency in our ablation studies, we utilize a $1$\,M representative subset of this collection.

Our exploration began by directly mixing VGGSound (Sparse label) with the Open-source T2A data (Fine-grained caption). However, this configuration consistently led to training divergence, where the loss would initially decrease but eventually spike. We attribute this failure to a severe semantic mismatch between the two text styles. Without a visual modality in the T2A samples to act as a grounding bridge, the text encoder embeddings for sparse labels and descriptive captions occupy disparate regions of the latent space, preventing the model from establishing a consistent audio-conditioning mapping.

To address this, we utilized Qwen3.0-Omni (30B)~\citep{Qwen3-Omni} to generate dense audio-visual descriptions for the VGGSound dataset, rephrasing them to align with the description style of the Open-source T2A data. This ``Dense'' VGGSound variant successfully stabilized the training when mixed with T2A data. However, as shown in Table~\ref{tab:app_datamix}, the resulting performance was inferior to the baseline trained solely on VGGSound (DeSync $0.52 \rightarrow 0.57$, IB $0.29 \rightarrow 0.26$). This suggests that while semantic alignment in text can prevent divergence, audio-only data cannot substitute for the rich structural information provided by paired video-audio samples in raw-space generation.

Consequently, we introduced the $1$\,M Media data subset to provide explicit visual supervision. Through our experiment, mixing VGGSound (Sparse label) with Media data (Fine-grained caption) converges reliably---unlike the T2A mixture---despite the disparity in text granularity. This confirms that the presence of the visual modality in both datasets allows the model to align concepts across different text styles by using visual features as a common semantic anchor.

\begin{table}[!t]
\caption{\textbf{Effect of data mixture and caption granularity on VT2A generation} ( WavFlow-M-16k, 1M training data scale, evaluated on VGGSound-Val). ``Open-source T2A'' refers to a mixture of FreeSound and AudioCaps. \textbf{Note:} Dense captions are applied to the VGGSound training set only; the VGGSound validation set utilizes native sparse labels for evaluation.}
\centering
\small
\label{tab:app_datamix}
\vspace{0.05in}
\renewcommand{\arraystretch}{1.1}
\setlength{\tabcolsep}{2.5pt}
\begin{tabular}{llcccccc}
\toprule
Data Mixture & VGG Caption
& FD$_{\text{PaSST}}\downarrow$
& FD$_{\text{PANNs}}\downarrow$
& KL$_{\text{PANNs}}\downarrow$
& IS$_{\text{PANNs}}\uparrow$
& IB$\uparrow$
& DeSync$\downarrow$ \\
\midrule
VGGSound (200K)             & Sparse label & 141.64 & 11.16 & 1.26 & 12.83 & 0.29 & 0.52 \\
VGGSound + Open-source T2A  & Sparse label & \multicolumn{6}{c}{\emph{training diverged (semantic mismatch)}} \\
VGGSound + Media data (1M)   & Sparse label & \textbf{121.09} & \textbf{9.58} & \textbf{1.20} & 16.42 & \textbf{0.33} & \textbf{0.47} \\
\midrule
VGGSound + Open-source T2A  & Dense  & 139.23 & 12.97 & 1.28 & 12.59 & 0.26 & 0.57 \\
VGGSound + Media data (1M)   & Dense  & 125.52 & 11.20 & 1.35 & \textbf{17.05} & \textbf{0.33} & 0.49 \\
\bottomrule
\end{tabular}
\end{table}

Finally, we compared the impact of caption quality within this stabilized mixture by evaluating VGGSound (Sparse label) + Media data against VGGSound (Dense) + Media data. The Dense variant yields higher IS$_{\text{PANNs}}$ ($17.05$ vs. $16.42$), indicating that fine-grained training descriptions help the model learn more diverse and complex audio-visual semantic mappings that generalize even when evaluation captions remain sparse. However, the Sparse label variant achieves superior FD and DeSync scores (e.g., FD$_{\text{PaSST}}$ of $121.09$ vs. $125.52$) due to the better consistency between the training and evaluation text distributions. Given its superior distributional fidelity and temporal synchronization, we select \textbf{VGGSound (Sparse label) + Media data} as our final training mixture.

\section{Waveform Patchify Configuration}
\label{app:tokenization}

\begin{figure}[H]
\centering
\includegraphics[width=0.85\linewidth]{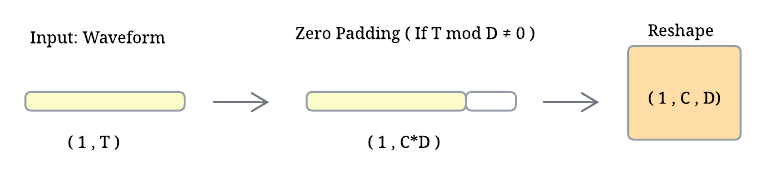}
\vspace{-0.1in}
\caption{\textbf{Waveform patchify illustration.} A 1D waveform is reshaped into a 2D token grid of shape $C \times D$. Zero-padding is naturally applied to handle arbitrary waveform lengths, which is then removed during unpatchify to recover the original $T$ samples.}
\label{fig:app_wavtok}
\end{figure}

As discussed in Section~\ref{para:method_patchify} and Section~\ref{para:patchify}, we reshape an $8$-second, $16$\,kHz waveform ($T=128{,}000$ samples) into a $C \times D$ token grid. As illustrated in Figure~\ref{fig:app_wavtok}, for audio samples where $T$ is not exactly divisible by $D$, we apply zero-padding to the waveform prior to patching; this padding is subsequently truncated during the unpatchify process to restore the original waveform length $T$.

We sweep the patch dimension $D$ from $512$ down to $160$ to determine the optimal granularity for waveform modeling (Table~\ref{tab:app_tokenization}). This sweep includes specific configurations, such as $C=576$ ($192 \times 3$) and $C=768$ ($192 \times 4$), designed to align the audio token count with the Synchformer feature length ($192$ tokens) to test if such explicit choice benefits temporal alignment.

\vspace{-0.10in}

\begin{table}[H]
\caption{\textbf{Ablation on patchify granularity} (WavFlow-M-16k, 3M training data scale, evaluated on VGGSound-Val). All configurations target an $8$\,s waveform ($T=128{,}000$).}
\centering
\small
\label{tab:app_tokenization}
\vspace{0.05in}
\renewcommand{\arraystretch}{1.1}
\setlength{\tabcolsep}{4pt}
\begin{tabular}{lccccccc}
\toprule
Patchify ($C \times D$)
& Token dur.\ (ms)
& FD$_{\text{PaSST}}\downarrow$
& FD$_{\text{PANNs}}\downarrow$
& KL$_{\text{PANNs}}\downarrow$
& IS$_{\text{PANNs}}\uparrow$
& IB$\uparrow$
& DeSync$\downarrow$ \\
\midrule
$250 \times 512$  & 32.0  & 139.45 & 11.93 & 1.41 & 16.14 & 0.312 & 0.48 \\
$500 \times 256$  & 16.0  & 124.02 &  9.84 & 1.18 & 16.72 & 0.335 & 0.48 \\
$576 \times 224$  & 13.9  & 125.69 & 10.02 & 1.20 & 16.92 & 0.339 & 0.45 \\
$640 \times 200$  & 12.5  & \textbf{120.88} & 10.02 & \textbf{1.16} & \textbf{17.20} & 0.334 & \textbf{0.45} \\
$768 \times 168$  & 10.4  & 124.05 &  9.91 & 1.18 & 17.00 & 0.338 & 0.47 \\
$800 \times 160$  & 10.0  & 121.83 &  \textbf{9.90} & 1.22 & 17.06 & \textbf{0.340} & 0.46 \\
\bottomrule
\end{tabular}
\end{table}
\vspace{-0.10in}

The results in Table~\ref{tab:app_tokenization} show that the coarsest configuration ($D=512$, $32$\,ms per token) performs significantly worse, indicating that the model requires finer granularity to capture waveform details. Once the patch dimension $D$ is reduced below $256$ ($16$\,ms), the generative performance and synchronization metrics stabilize at a high level.

Notably, we found that configurations specifically designed for sync-alignment (e.g., $C=576$ or $768$) did not yield substantial improvements over other fine-grained settings. This suggests that the model is robust to various token counts as long as the temporal resolution is sufficient. Considering the balance between generative quality, computational efficiency, and synchronization, we select \textbf{$640 \times 200$} as the default configuration for our $16$\,kHz experiments.

\section{Effect of Noise-Level Shift}
\label{app:noiseshift}

We evaluate the impact of noise-level shifting on the VGGSound test set. Prior work has shown that increasing the noise level during training can improve generation performance, particularly for high-resolution images~\citep{li2026basicsdenoise, hoogeboom2023simple}. Motivated by this finding, we investigate whether noise shift similarly benefits waveform-space generation.

We parameterize noise shift as $t_s = t / (t + s \cdot (1 - t))$ where $s$ is the shift factor, which biases training toward higher noise levels and reduces the effective signal-to-noise ratio by a factor of $s^2$.

\begin{table}[H]
\caption{\textbf{Effect of noise shift on VT2A generation} (WavFlow-M-16k, 1M training data scale, evaluated on VGGSound-Test). All configurations target an $8$\,s waveform ($T=128{,}000$).}
\centering
\small
\label{tab:app_noiseshift}
\vspace{0.05in}
\renewcommand{\arraystretch}{1.1}
\setlength{\tabcolsep}{4pt}
\begin{tabular}{lcccccc}
\toprule
Noise shift $s$
& FD$_{\text{PaSST}}\downarrow$
& FD$_{\text{PANNs}}\downarrow$
& KL$_{\text{PANNs}}\downarrow$
& IS$_{\text{PANNs}}\uparrow$
& IB$\uparrow$
& DeSync$\downarrow$ \\
\midrule
1.0 & \textbf{63.05} & \textbf{6.21} & 1.76 & \textbf{15.58} & \textbf{0.28} & \textbf{0.50} \\
3.0 & 73.17 & 7.52 & \textbf{1.72} & 13.64 & 0.26 & 0.50 \\
5.0 & 92.21 & 10.25 & 1.80 & 11.81 & 0.23 & 0.55 \\
\bottomrule
\end{tabular}
\end{table}

As shown in Table~\ref{tab:app_noiseshift}, unlike in image generation, noise shift provides little benefit for waveform-space modeling and progressively degrades all metrics as $s$ increases, with FD$_{\text{PaSST}}$ rising from $63.05$ ($s=1.0$) to $73.17$ ($s=3.0$) and then to $92.21$ ($s=5.0$). We attribute this to the fundamental difference in signal characteristics: image pixels span a broad value range where high-noise training helps the model capture low-frequency global structure, whereas audio waveforms have inherently low information density even after our preprocessing. In this regime, increasing the noise level makes the already weak audio signal even harder to recover, degrading rather than improving learning. We therefore adopt $s=1.0$ (no shift) as our default.

\section{Inference Hyperparameters}
\label{app:inference}

We ablate two key inference-time hyperparameters on VT2A (VGGSound validation set): the classifier-free guidance (CFG) strength and the number of ODE integration steps.

\begin{table}[!t]
\caption{\textbf{Effect of inference hyperparameters on VT2A} (WavFlow-M-16k, 1M training data scale, evaluated on VGGSound-Val). All configurations target an $8$\,s waveform ($T=128{,}000$).}
\centering
\small
\label{tab:ablation_inference}
\vspace{0.05in}
\renewcommand{\arraystretch}{1.1}
\setlength{\tabcolsep}{4pt}
\begin{tabular}{lcccccc}
\toprule
Setting
& FD$_{\text{PaSST}}\downarrow$
& FD$_{\text{PANNs}}\downarrow$
& KL$_{\text{PANNs}}\downarrow$
& IS$_{\text{PANNs}}\uparrow$
& IB$\uparrow$
& DeSync$\downarrow$\\
\midrule
\multicolumn{7}{l}{\textit{CFG strength (50 steps)}} \\
1.0  & 145.53 & 16.34 & 1.54 & 9.86  & 0.26 & 0.71 \\
2.5  & \textbf{108.19} & \textbf{9.40} & 1.26 & 15.44 & 0.32 & 0.51 \\
4.5  & 121.09 & 9.58 & \textbf{1.20} & \textbf{16.42} & \textbf{0.33} & 0.47 \\
7.0  & 142.00 & 9.92 & 1.22 & 16.19 & 0.32 & \textbf{0.46} \\
\midrule
\multicolumn{7}{l}{\textit{ODE steps (CFG=4.5)}} \\
10   & 118.90 & 10.73 & 1.30 & 14.93 & 0.31 & 0.47 \\
25   & \textbf{117.63} & 9.78 & 1.23 & 16.19 & 0.32 & 0.47 \\
50   & 121.09 & 9.58 & \textbf{1.20} & \textbf{16.42} & \textbf{0.33} & 0.47 \\
100  & 124.53 & \textbf{9.41} & \textbf{1.20} & 16.39 & \textbf{0.33} & 0.47 \\
\bottomrule
\end{tabular}
\end{table}

As shown in Table~\ref{tab:ablation_inference}, CFG strength has a pronounced effect on generation quality. Without sufficient guidance (CFG\,=\,1.0), the model produces low-fidelity outputs with poor semantic alignment (IB: 0.26) and severe temporal artifacts (DeSync: 0.71). Increasing CFG to 2.5 yields the best distributional fidelity (FD$_{\text{PaSST}}$: 108.19), while CFG\,=\,4.5 achieves the highest per-sample quality (IS: 16.42, IB: 0.33). Beyond 4.5, further increasing guidance to 7.0 causes FD$_{\text{PaSST}}$ to degrade sharply, indicating reduced output diversity without meaningful quality gains.

For ODE steps, the transition from 10 to 25 steps brings a significant quality improvement, and going from 25 to 50 steps yields further gains in IS ($16.19 \to 16.42$) and IB ($0.32 \to 0.33$). Beyond 50 steps, metrics plateau entirely, with 100 steps providing no additional benefit. We therefore adopt CFG\,=\,4.5 with 50 ODE steps as our default inference configuration, which offers the best generation quality before diminishing returns set in.

\section{Evaluation on MovieGen-Audio-Bench}
\label{app:moviegen_bench}

To further evaluate the generalization of \system, we conduct additional experiments on the MovieGen-Audio-Bench~\citep{polyak2025moviegencastmedia}. This benchmark is particularly challenging as it consists of AI-generated videos rather than real-world recordings. Performing audio synthesis for such synthetic content requires the model to establish a fundamental understanding of audio-visual correlation.

Since this benchmark does not provide ground-truth audio, we follow the protocol in~\citep{polyak2025moviegencastmedia} and report reference-free metrics: Inception Score (IS) for audio quality, CLAP and IB-score for semantic alignment, and DeSync for temporal synchronization. Table~\ref{tab:app_moviegen_bench} presents the comparison between \system, MMAudio, and MovieGen.

\begin{table}[H]
\caption{\textbf{Evaluation on MovieGen-Audio-Bench.} Best results are in \textbf{bold}, and second-best are \underline{underlined}. Reference-free metrics are used as no ground-truth audio is available for this benchmark.}
\centering
\small
\label{tab:app_moviegen_bench}
\vspace{0.05in}
\renewcommand{\arraystretch}{1.1}
\setlength{\tabcolsep}{6pt}
\begin{tabular}{lcccccc}
\toprule
Method & Params & Training Data & IS$\uparrow$ & CLAP$\uparrow$ & IB-score$\uparrow$ & DeSync$\downarrow$ \\
\midrule
WavFlow (ours) & 1.03B & $\sim$ 11.1K h & \textbf{8.95} & \underline{0.28} & 0.24 & \textbf{0.77} \\
MMAudio        & 1.03B & $\sim$ 8.2K h  & 8.40 & \underline{0.28} & \underline{0.27} & \textbf{0.77} \\
MovieGen       & 13B   & $\sim$ 1,000K h & \underline{8.89} & \textbf{0.29} & \textbf{0.36} & \underline{1.00} \\
\bottomrule
\end{tabular}
\end{table}
Experimental results show that \system generalizes effectively to synthetic visual content. Notably, \system achieves an Inception Score (IS) of 8.95 and a DeSync score of 0.77, outperforming or matching existing latent-based models in audio quality and temporal synchronization. While MovieGen maintains an advantage in semantic richness (IB-score), likely due to its significantly larger model capacity and training scale, \system remains highly competitive. These results suggest that direct raw-waveform synthesis, without the aid of a pretrained VAE, achieves performance comparable to state-of-the-art latent-based paradigms across both real and synthetic benchmarks.

\begin{figure}[H]
\centering
\includegraphics[width=0.9\linewidth]{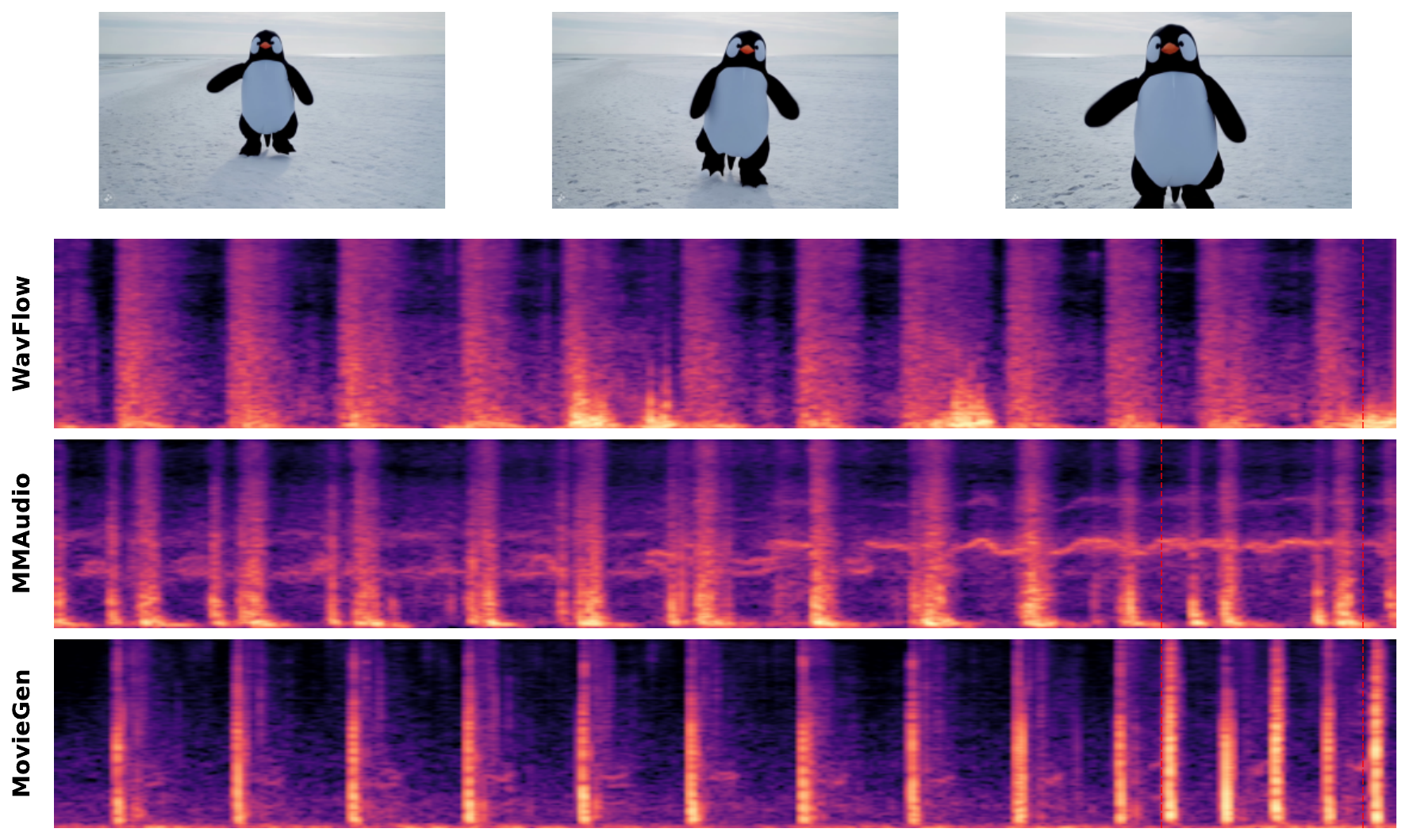}
\vspace{-0.1in}
\caption{\textbf{Spectrogram comparison on the "Penguin Walking" scenario.} All evaluated models demonstrate precise temporal synchronization. \system exhibits sharper and more vertical energy pulses, reflecting its ability to preserve fine-grained acoustic transients through direct raw-waveform modeling.}
\label{fig:qie}
\end{figure}

\begin{figure}[H]
\centering
\includegraphics[width=0.95\linewidth]{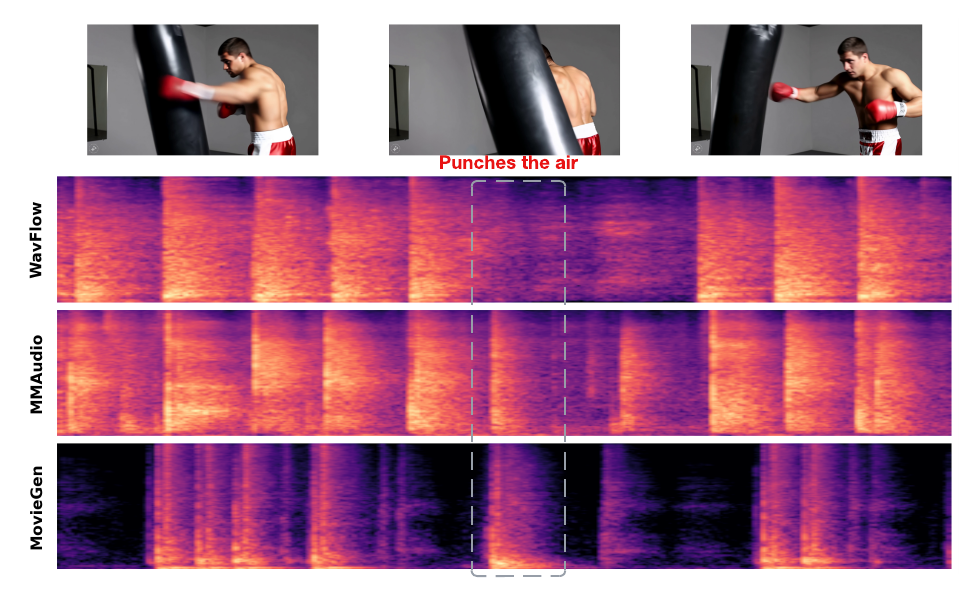}
\vspace{-0.1in}
\caption{\textbf{Spectrogram comparison on the "Boxing" scenario.} Acoustically, both \system and MMAudio achieve precise synchronization, while MovieGen exhibits desynchronization. Crucially, during the air-punching segment (no bag contact), \system correctly omits the impact sound, whereas MMAudio and MovieGen erroneously synthesize strike energy. This highlights \system's precision in discerning subtle visual nuances for accurate raw-waveform synthesis.}
\label{fig:punch}
\end{figure}

\begin{figure}[H]
\centering
\includegraphics[width=0.9\linewidth]{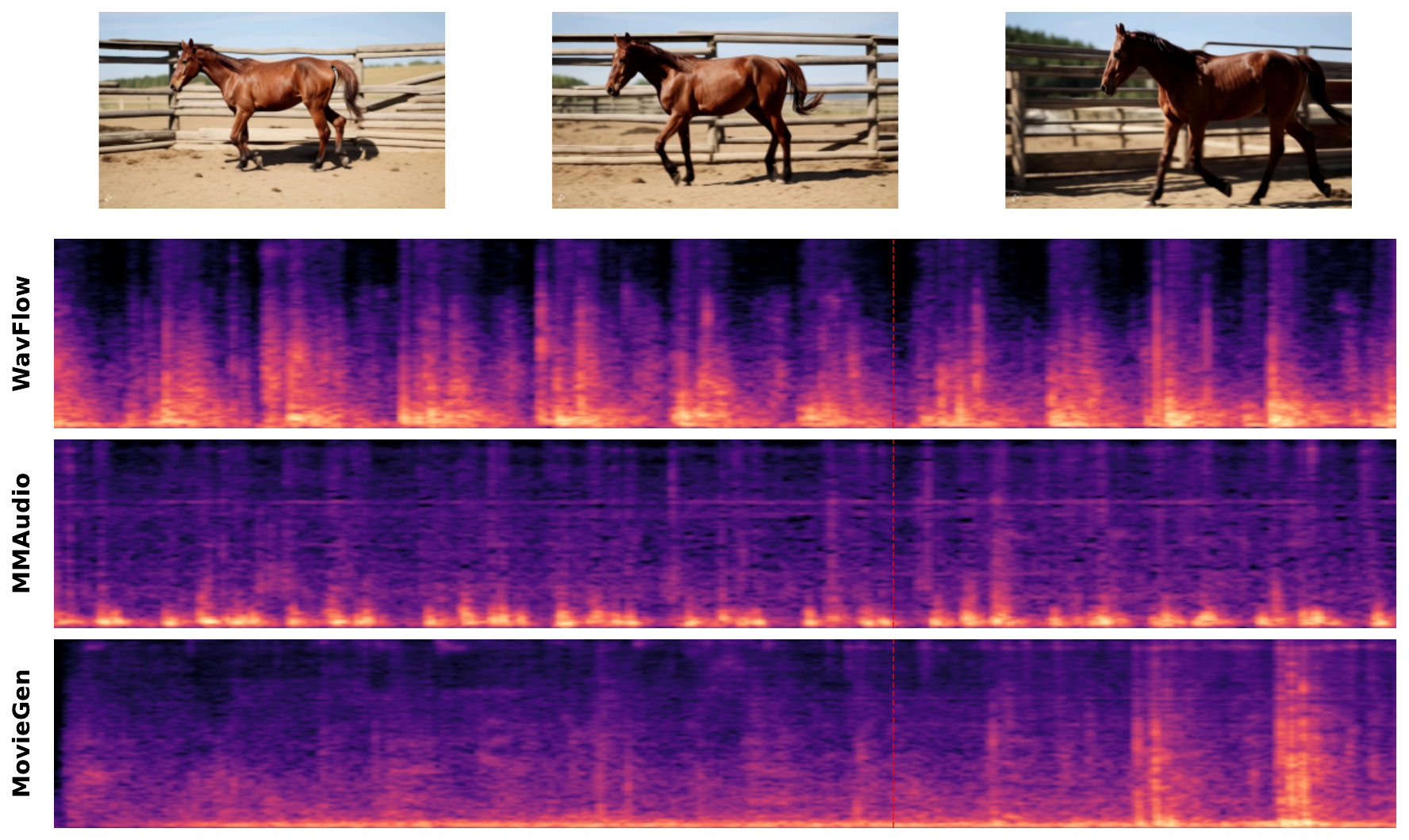}
\vspace{-0.1in}
\caption{\textbf{Spectrogram comparison on the "Horse Trotting" scenario.} In this scenario featuring consistent, rhythmic hoofbeats, \system achieves precise temporal synchronization, characterized by distinct and sharp vertical energy pulses that correspond to each footfall. While MMAudio maintains some rhythmic alignment, its spectral pulses are noticeably less defined and lack the transient sharpness seen in our model. In contrast, MovieGen fails to establish a clear temporal relationship with the visual motion.}
\label{fig:horse}
\end{figure}

To qualitatively evaluate temporal synchronization, we provide spectrogram visualizations of samples from the MovieGen-Audio-Bench in Figures~\ref{fig:qie} to \ref{fig:horse}. Detailed analyses for each case are provided in the respective figure captions.

\end{document}